\documentclass[11pt]{article}
\usepackage{epstopdf}                                                          
\usepackage[a4paper,hmarginratio=1:1,vmarginratio=2:3,totalwidth=15.3cm,totalheight=22.56cm]{geometry}
\usepackage{bm,epstopdf,epsfig,amsmath,amssymb,amsfonts,colordvi,latexsym,wrapfig,comment,cancel,verbatim,slashed}
\usepackage{epsfig,bbm}
\usepackage{graphicx,graphics}
\usepackage[font=md,captionskip=8pt]{subfig}
\usepackage[usenames,dvipsnames]{color}
\usepackage[noadjust]{cite}
\usepackage{xcolor} 
\usepackage[utf8]{inputenc}
\usepackage{setspace}

\usepackage[utf8]{inputenc}

\newcommand{\q}{\alpha}

\newcommand{\sg}{\sqrt{g}}    
\newcommand{\sgh}{\sqrt{\hat g}}
\newcommand{\w}{\omega}

\newcommand{\tGamma}{\tilde\Gamma}

\allowdisplaybreaks

\newcommand{\cL}{{\cal L}}

\newcommand{\cO}{{\cal O}}

\newcommand{\ra}{\rightarrow}

\newcommand{\be}{\begin{equation}}
\newcommand{\ee}{\end{equation}}
\newcommand{\bea}{\begin{eqnarray}}
\newcommand{\eea}{\end{eqnarray}}

\newcommand{\baa}{\begin{array}}
\newcommand{\eaa}{\end{array}}

\long\def\symbolfootnote[#1]#2{\begingroup
\def\thefootnote{\fnsymbol{footnote}}\footnote[#1]{#2}\endgroup}

\begin{document} 
\begin{flushright}
  % \today\\
\end{flushright}
\bigskip\medskip
\thispagestyle{empty}
\vspace{2cm}

\begin{center}
\vspace{0.5cm}

{\Large \bf  Non-metric geometry  as the origin of mass
  \bigskip

  in gauge theories of scale invariance}

 \vspace{1.5cm}
 
 {\bf D. M. Ghilencea}
 \symbolfootnote[1]{E-mail: dumitru.ghilencea@cern.ch}
 
\bigskip

{\small Department of Theoretical Physics, National Institute of Physics
 \smallskip 

 and  Nuclear Engineering (IFIN), Bucharest, 077125 Romania}
\end{center}

\medskip

\begin{abstract}
  \begin{spacing}{0.99}
    \noindent
We discuss  gauge theories of scale invariance beyond the Standard Model (SM) and
Einstein gravity. A consequence of gauging this symmetry is that their
underlying 4D geometry is non-metric ($\nabla_\mu g_{\alpha\beta}\!\not=\!0$).
Examples of such theories are  Weyl's {\it original} quadratic gravity theory and its Palatini
version. These theories have spontaneous breaking of the gauged scale symmetry to
Einstein gravity. All mass scales have a  geometric origin:
the Planck scale ($M_p$),  cosmological constant ($\Lambda$) and the
mass  of the Weyl gauge boson  ($\omega_\mu$) of scale symmetry are proportional
to a  scalar field vev that has an origin in the (geometric) $\tilde R^2$ term in the action.
With $\omega_\mu$ of non-metric geometry origin, the  SM  Higgs field  also has a similar
origin,   generated  by Weyl boson fusion in the early Universe.
This appears as a microscopic realisation
of ``matter creation from geometry'' discussed in the thermodynamics of open systems
applied to cosmology. Unlike in  local scale invariant theories (no $\omega_\mu$ present)
with an  underlying  pseudo-Riemannian geometry, in our case:
{\bf 1)}  there are no ghosts and no additional fields beyond the SM and
 underlying Weyl or Palatini geometry,
{\bf 2)} the cosmological constant is  positive and is small because gravity  is  weak,
{\bf 3)} the  Weyl or Palatini connection shares  the Weyl (gauge) symmetry of the action, and:
{\bf 4)} there exists  a  non-trivial, conserved Weyl current of this symmetry.
An intuitive picture of  non-metricity  and its relation to mass generation
is also provided from a  solid state physics perspective where it is  common and is
associated with point defects (metric anomalies) of the crystalline structure.
\end{spacing}
\end{abstract}

\newpage

\section{Introduction}

Scale symmetry may play a role in physics beyond the Standard Model (SM) and Einstein gravity.
This is suggested by the fact that    the  SM with a vanishing Higgs mass (parameter)
is scale invariant \cite{Bardeen}. Moreover, at high energies or in the early Universe, the
states of the SM are essentially massless and the theory can have a scale symmetry
(global, local or gauged scale symmetry). But Einstein gravity breaks such symmetry,
hence one can attempt to generate it as  a spontaneously   broken phase  of a theory with a
local or gauged scale  symmetry (regarding global scale symmetry, it does not survive
black hole physics \cite{Kallosh}).

Since gravity ``is''  geometry,  the first question is what  the
underlying 4D space-time  geometry of a theory beyond  Einstein gravity and SM  is.
If the action is  locally scale invariant,  one could expect
that this should also be, for consistency, a symmetry of the underlying
geometry  i.e. of the connection. But the  (pseudo)-Riemannian geometry and its
Levi-Civita connection are not (Weyl) locally scale invariant. One may then seek 
an alternative geometry whose  connection has the space-time symmetry of the action.
A stronger motivation to do so is the gauge principle: similarly to the SM as a
(quantum) gauge theory, we seek  a {\it gauge}  theory of scale invariance that recovers Einstein
gravity in its broken phase.

This principle leads us  to consider the Weyl conformal geometry \cite{Weyl1,Weyl2,Weyl3}
because Weyl connection does have a gauged scale symmetry (also known as {\it Weyl gauge symmetry}).
We also consider the  Palatini approach to gravity \cite{EP} 
where the offshell connection, being independent of the metric, has this symmetry, too.
In a gauged scale invariant theory its underlying Weyl  geometry is {\it non-metric - by definition
this means that} $\tilde\nabla_\mu g_{\alpha\beta}\!\not=\!0$;
the Palatini version with such symmetry is also non-metric \cite{Palatini1}.
This differs  from other theories beyond SM and Einstein gravity that assume 
that the underlying  geometry is a metric (pseudo)Riemannian geometry  
($\nabla_\mu g_{\alpha\beta}\!=\!0$),
 {\it e.g.\,}conformal gravity \cite{Mannheim},  supergravity \cite{vanProeyen}
 or strings embedding \cite{Z6}.  In our case,
 non-metricity is present because the  gauge field of scale transformations ($\w_\mu$) is
  dynamical ($F_{\mu\nu}\not=0$) i.e. physical,
 as actually  expected in a  true gauge theory; this is not true in
 conformal gravity \cite{Kaku} (also \cite{Wheeler,Freund}) which is then
 a {\it metric} theory. Briefly,  non-metricity is  due to the
 dynamics\footnote{A dynamical $\omega_\mu$
   agrees with the view that at fundamental level gravity\,may\,be\,a
   theory\,of connections\,\cite{ST2}.}
 of $\omega_\mu$  and, interestingly,  the same is true for the  mass generation in such
 theory because $F_{\mu\nu}$ is the {\it length} curvature tensor.
Hence, non-metricity and mass generation are  related, as we discuss.

Weyl geometry was  criticised for its non-metricity \cite{Weyl1} but remained of interest  \cite{Scholz}
even though  it failed to describe gravity ``plus'' electromagnetism as Weyl initially intended
\cite{Weyl1,Weyl2,Weyl3}. Actually,  the  same  {\it original} theory of Weyl,
which is a {\it quadratic} gravity  theory of action $\sqrt{g}\,(\tilde R^2\!-\!F_{\mu\nu}^2)$,
defined by Weyl geometry,  is a realistic gauge theory of scale symmetry
as  first shown in  \cite{Ghilen1}: it has a spontaneous breaking
(Stueckelberg mechanism) to Einstein gravity 
plus a Proca action of the Weyl gauge boson of scale symmetry. This boson\footnote{
  which Weyl unfortunately (and wrongly) attempted to identify with the massless, real photon.}
 is thus   a {\it  massive} gauge field of dilatations that decouples at
 a high scale (mass $\!\propto\! M_p$), below which  a (metric) (pseudo)-Riemannian
 geometry and Einstein gravity are found \cite{Ghilen1,SMW}.
 So Weyl geometry  and its gauged scale invariance give a UV completion of 
 (pseudo)-Riemannian geometry and Einstein gravity.
 We thus  have a gauge theory embedding of Einstein gravity
 - the latter is simply a ``low energy''  {\it broken phase} of Weyl's original  gravity theory.

 In theories  with a scale symmetry of the (canonical) Lagrangian, dimensionful
 couplings are forbidden. Then  their mass scales e.g. Planck scale ($M_p$),
 cosmological constant ($\Lambda$),  are generated by vacuum expectation values (vev)
of some {\it additional}  scalar field(s) beyond the SM Higgs. 
These  extra fields are  added  {\it ad-hoc}  as a ``patch up'' solution and
often are {\it  ghosts}. Further, sometimes the local scale symmetry used is  a  ``fake symmetry''
since  its associated current is trivial. We want to avoid all these
issues (see Section~{\bf\ref{metric}}).
Finally, with gravity  related to the underlying geometry,
could $M_p$ and $\Lambda$ have a  common,  {\it geometric} origin?

In this work  we review  the role of non-metricity
in solving these problems, based on our results in \cite{Ghilen1,SMW,Palatini1}.
We compare  (Section~{\bf\ref{MG}}) the special, metric case (where $\w_\mu$ is not dynamical)
to the non-metric case of  Weyl {\it quadratic} gravity \cite{Ghilen1,SMW} and of its Palatini
version \cite{Palatini1}\footnote{Prior to \cite{Ghilen1,SMW,Palatini1}
  realistic non-metric  theories were  linear-only
  in $R$ and had extra scalars e.g.\cite{Smolin,Moffat,Guendelman}.}.
We show how (non-metric) geometry  generates  {\it all mass scales} in both Weyl
and Palatini cases, {\it in the absence of matter,} while avoiding the aforementioned issues.
These results remain valid if matter is included e.g. the SM.
We shall see  (Section~{\bf\ref{pheno}}) how non-metric geometry (in essence $\omega_\mu$)
can be responsible for  generating the SM Higgs field: this gives a  microscopic picture of
``matter creation from geometry'' usually discussed in the  thermodynamics of
open systems applied to cosmology \cite{P1,P2,P4}.
We also discuss the absence of a  ``second clock effect'' (the
initial Einstein's critique \cite{Weyl1} of non-metricity) in a spontaneously broken gauge
theory of scale invariance such as Weyl or Palatini theory. Given its important role
here, we also provide a more intuitive picture of non-metricity  from the
solid state physics perspective where it is associated with point defects
of the crystalline structure. After Conclusions (Section~{\bf\ref{conc}}),
Appendix {\bf A} and {\bf B} present technical details.

\section{A ``metric'' example}\label{metric}

To detail the above  ideas, consider first the (pseudo-)Riemannian
geometry and\footnote{Our convention  is  $g_{\mu\nu}=(+,-,-,-)$, $g=\vert\det g_{\mu\nu}\vert$
  while the curvature tensors are defined as in  \cite{book}.}
%\medskip
\bea\label{E}
\cL_E=-\frac12\,\sqrt{g}\, (M^2_p R +2\,M^2_p \Lambda)
\eea
%
%\medskip\noindent
where $M_p$ is the Planck scale and $\Lambda$ is the cosmological constant.
One can regard $\cL_E$ as a spontaneously broken phase of a locally scale invariant
action; this symmetry is defined by invariance under (i) below,
extended by (ii) if real scalars ($\phi$) or fermions
($\psi$) are  present
\bea\label{WS}
\textrm{(i)} &\quad& \hat g_{\mu\nu}=\Sigma^q %\Omega^2
\,g_{\mu\nu},\qquad 
\sqrt{\hat g}=\Sigma^{2 q} \sqrt{g},
\nonumber\\[5pt]
\text{(ii)} &\quad & \hat \phi = \Sigma^{-q/2} \phi, \qquad \hat\psi=\Sigma^{-3q/4}\,\psi,
\qquad\qquad\quad (q=1).
\eea
where $g=\vert\det g_{\mu\nu}\vert$, $\Sigma(x)>0$ and we set the charge
$q\!=\!1$ without loss of generality\footnote{The case of arbitrary $q$ in 
  transformations
  (\ref{WS}) and (\ref{WGS}) is recovered by  replacing  $\alpha\ra q\alpha$ in the results.}.
Consider  implementing this symmetry\footnote{
For  some interesting models  beyond the SM  with this symmetry see e.g.
\cite{Bars0,Bars1,Bars2,Kallosh-2,tH0,tH3}.} using \cite{Brans,Jordan},
so\,one adds `by hand' a scalar $\phi$, then
% \medskip
\bea\label{RM}
\cL_E=\sqrt{g}\,\Big\{\frac{-1}{2} 
\Big[ \frac{1}{6}\,\phi^2 \,R+(\partial_\mu\phi)^2\Big]
\pm\lambda \phi^4\Big\},
\eea
% \medskip\noindent
is invariant under ($\ref{WS}$).
By a formal transformation (\ref{WS}), with $\Sigma=\phi^2/\langle \phi\rangle^2$,
one fixes the gauge of this symmetry i.e. fixes
$\hat\phi$ to a constant vev,  assumed to exist, $\langle\phi^2\rangle=6 M_p^2$, so
gauge fixing confirms a dynamical breaking (not vice-versa).
Then one generates the first term in  (\ref{E}) and $\phi$ decouples.
Regarding the second term in (\ref{E}), this can be obtained only if one is adding ``by hand''
a $\phi^4$ term to (\ref{RM}) {\it and} with the right sign!

For our later  comparison to  ``non-metric'' cases, notice the following
(see also \cite{Quiros1,Quiros2,Ohanian}):

\noindent
{\bf a)\,} Symmetry (\ref{WS}) enforces the
sign of the kinetic term to be  negative so $\phi$ is a ghost.
It is  common in conformal/superconformal models that
$\phi$, which acts as a compensator rather than a physical field,
has a kinetic term of negative sign%
\footnote{
  This has additional implications e.g. for
  the scalar potential  in 4D N=1 supergravity \cite{vanProeyen}.}.
We would like to avoid a ghost in a classical action\footnote{
 Alternatively, if one changed the sign  of  $\cL_E$, then 
 $\langle\phi^2\rangle\!\!<\!0$,  $\Sigma\!<\!0$ but then
 $g_{\mu\nu}$ changes signature by (\ref{WS}).}.
Note also that the sign of $\Lambda\sim\langle\phi\rangle^2$ is {\it arbitrary}, not fixed by (\ref{WS}), (\ref{RM}).

\noindent
{\bf b)\,\,} In the absence of matter
the current $J^\mu$ associated to symmetry (\ref{WS})  is  {\it trivial} $J^\mu\!=\!0$;
this  raised concerns on the physical meaning of  symmetry (\ref{WS})
which was thus  called ``fake symmetry'' \cite{J1,J2}. We want to know
if a non-trivial current exists in  more general cases.

\noindent
{\bf c)\,\,}
$\phi$ is added ``by-hand'' to enforce  symmetry (\ref{WS}), as a ``compensator'',
so it is {\it  not related }  to the underlying geometry  of Einstein gravity
 emergent in the broken phase. So 
 the Planck scale  generated by $\langle\phi\rangle$ is not related to geometry.
 Can $M_p$, $\Lambda$  and $\phi$ have a geometric origin?

\noindent
{\bf d)\,\,} while  $\cL_E$  is invariant under (\ref{WS}),
the underlying geometry i.e. the  Levi-Civita connection is not!
Is it really consistent to have a space-time symmetry of an
action while its  underlying geometry (connection) does not have such symmetry?
Can we avoid this?  

In the following we shall see how the above issues  {\bf  a), b), c), d)} are
elegantly answered in the gauge theory of scale symmetry of  Weyl, based on Weyl geometry
or  in its  Palatini version. They ensure  a  geometric interpretation of this 
symmetry, something that is not so obvious in the  local scale  symmetry of
eq.(\ref{WS}) (for a discussion on this \cite{Bars0,Ohanian}).

\section{Non-metric geometry as the origin of mass}\label{MG}

\subsection{Weyl geometry: metricity vs non-metricity}\label{WMN}

\medskip\noindent
{\bf $\bullet$ Metric (integrable) case:  }

\medskip\noindent
Consider first   the case of Weyl conformal geometry\footnote{For a brief 
introduction to  Weyl conformal geometry and relevant formulae see Appendix A in  \cite{SMW}.}.
Weyl geometry is
defined by classes of equivalence $(g_{\alpha\beta}, \w_\mu$) of the metric ($g_{\alpha\beta}$)
and the Weyl gauge  field ($\w_\mu$),  related by the {\it Weyl gauge
symmetry} transformation. This symmetry is defined by transformation (\ref{WS}) {\it together with}
that of $\omega_\mu$ given by:
%\medskip
\bea\label{WGS}
\hat\w_\mu=\w_\mu -\frac{1}{\q}\, \partial_\mu\ln\Sigma, %\Omega^2
\eea

\medskip\noindent
with  $\alpha$ the Weyl gauge coupling. The {\it non-metricity} is defined below, by
the presence of $\w_\mu$:
\bea\label{nm}
\tilde\nabla_\mu g_{\alpha\beta}=-\alpha \,\w_\mu g_{\alpha\beta},
\eea

\smallskip\noindent
 $\tilde\nabla$ is defined by the connection $\tilde\Gamma$ of Weyl geometry, see eq.(\ref{sdsd}).
The solution is (eq.(\ref{tGammap}))
%\smallskip
\bea\label{tGamma}
\tilde \Gamma_{\mu\nu}^\lambda=
\Gamma_{\mu\nu}^\lambda+(1/2)\,\q \,\big[\delta_\mu^\lambda\,\, \w_\nu +\delta_\nu^\lambda\,\, \w_\mu
- g_{\mu\nu} \,\w^\lambda\big].
\eea

\smallskip\noindent
where $\Gamma$ is the Levi-Civita connection
$\Gamma^\alpha_{\mu\nu}(g)=(1/2) g^{\alpha\lambda} (\partial_\mu g_{\lambda\nu}+\partial_\nu g_{\lambda\mu}
-\partial_\lambda g_{\mu\nu})$. $\tilde \Gamma$ is {\it invariant} under combined
(\ref{WS}), (\ref{WGS}). If $\w_\mu$ decouples ($\w_\mu=0$) or is ``pure gauge'',
the theory is Weyl integrable and  metric.
Denote by $\tGamma_{\mu\nu}^{\nu}=\tGamma_\mu$,   $\Gamma_{\mu\nu}^{\nu}=\Gamma_\mu$, then 
%\medskip
\bea
\w_\mu=(1/2) \,(\tilde\Gamma_\mu-\Gamma_\mu)
\eea
$\w_\mu$
measures the deviation (of the trace) of the connection from the Levi-Civita connection.
Since $\w_\mu$ is part of the connection $\tilde \Gamma$,  it obviously  has a (non-metric) geometric origin.

The simplest gravity  action in Weyl  geometry, with symmetry (\ref{WS}), (\ref{WGS})  is
\bea\label{L1}
\cL_1=\sqrt{g}\,\frac{1}{4!\,\xi^2} \tilde R^2,\qquad \xi<1.
\eea
%
%\medskip\noindent
where  $\tilde R=R(\tilde \Gamma,g)$ is the scalar curvature of Weyl geometry,
defined by $\tilde \Gamma$ of (\ref{tGamma}) with the usual formulae.
 $\cL_1$ is invariant under (\ref{WS}), (\ref{WGS}). This is because
  $\tilde R=g^{\mu\nu} \tilde R_{\mu\nu}(\tilde\Gamma)$ where
 $\tilde \Gamma$ is invariant, hence under (\ref{WS}), (\ref{WGS}), $\hat{\tilde R}=\tilde R/\Sigma$
 and $\cL_1$ is invariant. One can then show
\medskip
\bea\label{tildeR}
\tilde R=R-3\,\q\,\nabla_\mu\w^\mu-\frac32 \q^2\,\w_\mu \,\w^\mu,
\eea
where the  rhs  is in a Riemannian notation, so
$\nabla_\mu\w^\lambda=\partial_\mu \w^\lambda+\Gamma^\lambda_{\mu\rho}\,\w^\rho$.
One can replace (\ref{tildeR})  in $\cL_1$. Since $\tilde R^2$ contains Riemannian
$R^2$, $\cL_1$ is  a higher derivative theory that propagates a
spin-zero mode (from  $R^2$), in addition to the graviton.
It is easy to ``unfold'' this higher derivative theory into a second order one
and extract this spin-zero mode
from $\tilde R^2$. To this purpose, replace
$\tilde R^2\ra - 2 \phi^2 \tilde R-\phi^4$ in $\cL_1$, where $\phi$ is a scalar field, to
obtain 
\bea\label{oo}
\cL_1=\sqrt{g}\,\frac{1}{4!\,\xi^2}\,\Big[- 2 \phi^2 \tilde R-\phi^4\Big].
\eea
The equation of motion of $\phi$ has
solution $\phi^2=-\tilde R$ which replaced in the action recovers  eq.(\ref{L1}),
so eqs.(\ref{L1}) and (\ref{oo}) are classically equivalent. Next, the  equation of motion of $\w_\mu$ is
\bea\label{omu}
\w_\mu=(1/\q)\, \partial_\mu\ln \phi^2,
\eea
so $\w_\mu$ is ``pure gauge'' and can be integrated out. Using this back in the action, then
\medskip
\bea\label{RMW}
\cL_1=\sqrt{g}\,\frac{1}{\xi^2}
\Big\{-
\frac{1}{2}\,
\Big[\,\frac16 \,\phi^2\,R + g^{\mu\nu} \partial_\mu \phi\,\partial_\nu \phi \Big]-
\frac{1}{4!}\,\phi^4\Big\}.
  \eea

  \medskip
 This  is called a Weyl {\it integrable} case since  $\omega_\mu$ is not dynamical:
 its field strength $F_{\mu\nu}=0$ due to (\ref{omu}). Hence, $\w_\mu$ is not a physical
field and then this theory is {\it not} a true gauge theory, however it is metric:
we apply (\ref{WGS}) with $\Sigma=\phi^2$, therefore  $\hat\w_\mu=0$, $\tilde\Gamma=\Gamma$,
$\tilde\nabla_\mu g_{\alpha\beta}=0$ so the connection is Levi-Civita and
the geometry becomes (pseudo)Riemannian.

  $\cL_1$ has similarities to the case of previous section, see text after eq.(\ref{RM}),
  but there are some good features  in the case here. 
  Firstly,  unlike in (\ref{RM}), here $\phi$ was {\it not} added ``ad-hoc'' but
  it came from the $\tilde R^2$ term in the action.
  Secondly, the last term in (\ref{RM}) has a definite sign.
  Assuming  that $\phi$ acquires  a vev (e.g. at quantum level, etc),
 by  applying (\ref{WS}) to (\ref{RMW}) with $\Sigma=\phi/\langle\phi\rangle$,
   (or formally  setting $\phi=\langle\phi\rangle$ in (\ref{RMW})),
  one   obtains eq.(\ref{E}) of Einstein action and a cosmological
  constant term, with a Planck mass $M_p^2=\langle \phi\rangle^2/(6\xi^2)$
  and $\Lambda=\langle\phi\rangle^2/4$.

 Briefly,  Weyl quadratic
 gravity in the {\it integrable case}  has certain advantages compared to the
 ``metric'' case in Section~\ref{metric}:  it generates Einstein gravity and
 it {\it explains} the origin 
  of both $M_P$ and $\Lambda$  as due to (a vev of)  $\phi$ which was {\it not added} by hand
  but it has a {\it geometric} origin in the
  $\tilde R^2$ term. The case here predicts a non-zero (positive) $\Lambda$,
   because both $\Lambda, M_p\propto\langle\phi\rangle^2$.
  This also suggests  a UV-IR connection in the physics associated to these two scales.

   However, similar to   Section~\ref{metric},    for action (\ref{RMW}) the
   current $J^\mu$ associated to  symmetry (\ref{WS}) is trivial $J^\mu\!=\!0$, 
   see \cite{J1,J2}, hence their conclusion that (\ref{WS}) is a fake symmetry.
   The  negative sign kinetic term in (\ref{RMW}) is also a concern
   in some cases. We want to avoid these two issues, but to  retain
   the good features found above. This is possible in the non-metric case.

\bigskip  \noindent
{\bf $\bullet$ Non-metric case:\,\,\,}

\medskip\noindent
The  above situation improves further if $\w_\mu$ has a kinetic term. 
  This brings us to the  original  Weyl gravity action
  \cite{Weyl1,Weyl2,Weyl3} which has a { gauged} scale symmetry.
  The action is
  \smallskip
  \bea\label{inA}
\cL_1^\prime=\sg\, \,\Big[\, \frac{1}{4!}\,\frac{1}{\xi^2}\,\tilde R^2
- \frac14\, F_{\mu\nu}^{\,2}\Big].
\eea

\smallskip\noindent
Here
$F_{\mu\nu}=\tilde\nabla_\mu\w_\nu-\tilde\nabla_\mu\w_\nu$ is the field strength of $\w_\mu$,
with $\tilde\nabla_\mu\w_\nu=\partial_\mu\w_\nu-\tilde\Gamma_{\mu\nu}^\rho\w_\rho$.
Since  $\tilde\Gamma_{\mu\nu}^\alpha\!=\!\tilde\Gamma_{\nu\mu}^\alpha$ is symmetric,
$F_{\mu\nu}=\partial_\mu\w_\nu-\partial_\nu\w_\mu$, just like in  flat space-time.

We ``linearise'' the $\tilde R^2$  term in (\ref{inA}) by replacing
$\tilde R^2\ra -2\phi^2 \tilde R -\phi^4$, where $\phi$ is a scalar field.
In the new action, the solution of the equation of motion of $\phi$ is
$\phi^2=-\tilde R$ which replaced in the action
recovers eq.(\ref{inA}), therefore we obtain a classically equivalent action.
Then we replace $\tilde R$ in terms of Riemannian $R$, eq.(\ref{tildeR}), to find 
\cite{Ghilen1,SMW}:
%\medskip
\bea\label{alt2}
\!\cL_1^\prime\!\!\!\!&=&\!\!\!
\sqrt g\,\Big\{
\frac{1}{4!\,\xi^2} \big[ - 2\phi^2 \,\tilde R-\phi^4\big] - \frac14 \,F_{\mu\nu}^2\,\Big\}
\\[3pt]
\!\!\!&=&\!\!\!
\!\sqrt g\,
\Big\{\frac{-1}{2\,\xi^2}\,
\Big[ \frac{\phi^2}{6} R
+(\partial_\mu\phi)^2
-\frac{\q}{2} \nabla_\mu (\w^\mu\phi^2)\Big]
-\frac{\phi^4}{4!\,\xi^2}
+\frac{\q^2}{8\,\xi^2} \phi^2 \Big[\w_\mu -\frac{1}{\q}\partial_\mu \ln\phi^2\Big]^2\!\!
-\frac{1}{4} F_{\mu\nu}^2\Big\}.\nonumber
\eea

\medskip
\noindent
From this Lagrangian  one finds (see Appendix A) that
\smallskip
\be\label{current}
J^\mu=-\q/(4\xi^2)\, g^{\mu\nu}\,(\partial_\nu -\alpha \,\w_\nu)\phi^2
\ee

\medskip\noindent
is a  {\it conserved current:} $\nabla_\mu J^\mu=0$ \cite{SMW,Ghilen1,Palatini1}.
The presence here  of a non-trivial conserved $J^\mu$
is a fundamental difference from the metric/integrable cases discussed above  and avoids the
criticisms of \cite{J1,J2}.
For the previous case of  Weyl integrable geometry of 
 eq.(\ref{RMW}), with $\omega_\mu$ of (\ref{omu})  one easily verifies that 
for action (\ref{RMW}) $J^\mu=0$ i.e.  the current  is trivial.

From current conservation equation  $\nabla_\mu J^\mu=0$ or from  eq.(\ref{opi}):
%\medskip
\bea\label{con}
\Box\phi^2 - \alpha\nabla^\rho \,(\omega_\rho\phi^2)=0, \qquad \Box=\nabla^\mu \nabla_\mu.
\eea
%
% \medskip\noindent
This shows that $\phi$ is  a {\it dynamical} field, as also seen from (\ref{RMW}),
 because $\tilde R^2$ contains the higher-derivative Riemannian $R^2$ 
that  propagates a spin-zero mode beyond graviton.
This  generalises 
a  conserved current   $\nabla^\mu (\nabla_\mu \phi^2)\!=\!0$ of global case \cite{Ga,Fe1,Fe2,Fe3,Fe4}
 recovered here for $\alpha\!=\!0$.

One would like to  ``fix the gauge'' of Weyl gauge symmetry. As in the previous metric and integrable cases,
assume that $\phi$ develops a vev e.g.  at the quantum level or as in  the global case
\cite{Ga,Fe1,Fe2,Fe3,Fe4}.
For example in a Friedmann-Robertson-Walker universe, with an ``isotropic''
  $\omega_\mu=(\omega_0(t), 0,0,0)$, if $\omega_0(t)\sim 1/\phi(t)^2$, then (\ref{con}) gives
  $\Box\phi^2\approx 0$ whose solution $\phi(t)$ evolves
  to a constant value (vev) at large $t$ \cite{Ga,Fe1,Fe2,Fe3,Fe4}, so $\phi\ra\langle\phi\rangle$.
  Then, on the ground state
  the  current conservation $\nabla_\mu J^\mu=0$ gives $\nabla^\rho\omega_\rho=0$,
specific to a massive Proca field $\w_\mu$. This  ``fixes the gauge''
 of symmetry (\ref{WS}), (\ref{WGS}). 
 At the level of the Lagrangian, this  gauge fixing is implemented by applying to $\cL_1^\prime$
transformation (\ref{WS}), (\ref{WGS}) with a special
$\Sigma=\phi^2/\langle\phi^2\rangle$; 
(or formally replace $\phi\ra \langle\phi\rangle$ in eq.(\ref{alt2})).
In terms of the   transformed fields (with a ``hat''),  $\cL_1^\prime$ becomes:
\medskip
\be
\label{EP}
\cL_1^\prime=\sgh \,\Big[- \frac12 M_p^2\hat R +    %\frac34 M_p^2\,\q^2\,
\frac12\, m_\w^2\, \hat\w_\mu \hat \w^\mu
-\Lambda\,M_p^2
-\frac{1}{4} \, \hat F_{\mu\nu}^2 \Big],
\ee
with 
\be\label{massesW}
M_p^2= \frac{\langle\phi^2\rangle}{6\,\xi^2},
\qquad
\Lambda = \frac14 \langle\phi^2\rangle,
\qquad
m_\w^2 
=\frac{\q^2\langle\phi^2\rangle}{4\,\xi^2}.
%= \frac32\, \q^2 M_p^2
\ee

\medskip\noindent
Eq.(\ref{EP}) is the Einstein-Proca Lagrangian for the Weyl vector \cite{Ghilen1,SMW},
in the Einstein gauge ("frame").  The Weyl gauge field
has absorbed the derivative of the field\footnote{See the
  second-last  term
  in the second line of eq.(\ref{alt2}). Here $\ln\phi$ has a shift symmetry eq.(\ref{WS}) and
  plays the role of a would-be Goldstone field  of the Weyl gauge symmetry. For a detailed
  discussion see \cite{Ghilen1,SMW,Palatini1}.}
$\partial_\mu(\ln\phi)$ via a Stueckelberg mechanism \cite{ST1},
with the total number of degrees of freedom (dof) conserved,
as expected for a spontaneous breaking:  the massless  $\w_\mu$  (dof=2) and
real, dynamical $\phi$ (dof=1) are replaced by a massive Proca field
$\w_\mu$ (dof=3, since $\nabla_\rho\omega^\rho=0$) \cite{Ghilen1,SMW}.
Notice that there is no ghost in (\ref{EP}).
The  mass of $\w_\mu$,  $m_\w\!\sim\!\q \, M_p$,  is close to $M_p$ unless
one is tuning $\q\!\ll\! 1$; hence, any (unwanted)  non-metricity effects  due to $\w_\mu$
are  suppressed by $m_\omega$.\footnote{Notice that one also  obtains that
  $\langle\phi^2\rangle =4\Lambda=-R=12 H^2$ in a FRW universe,
  so $\langle\phi^2\rangle=12 H^2$ \cite{SMW}.}.

Since $\w_\mu$ is massive, it decouples 
to leave in the broken phase below $m_\w$ the Einstein gravity and a {\it positive}
cosmological constant. At the same time, since $\w_\mu$ decouples from the action,
metricity {\it is restored} below $m_\w$: the connection $\tGamma$ of (\ref{tGamma}) becomes
Levi-Civita ($\Gamma$) and the geometry becomes Riemannian. 
Hence, Weyl geometry with its gauged scale invariance
 acts as a non-metric ultraviolet (UV)
completion (above $m_\w$)  of  Riemannian geometry.

\subsection{Palatini theories: metricity vs non-metricity}

\noindent
{\bf $\bullet$  Metric case:\,\,\,}

\medskip\noindent
Let us now consider  the Palatini approach
\cite{EP} to  quadratic gravity actions with a Weyl gauge symmetry, such as
eqs.(\ref{L1}), (\ref{inA}).
In this approach the connection is  independent of the metric and so it
is invariant under (\ref{WS}), (\ref{WGS}). The connection is then determined
from its equations of motion and this 
solution is used back in the initial action.
Hence, the underlying geometry (connection) and thus its  metricity or non-metricity
are determined by the  action and by its symmetries, as we shall see shortly:
if the theory is invariant under (\ref{WS})
the theory is metric; while if it is Weyl gauge invariant,  it is non-metric.

To begin with,  consider first an action  with symmetry  (\ref{WS})
in a  Palatini approach
\bea\label{L2}
\cL_2=\sqrt{g}\frac{1}{4! \,\xi^2}\,R(\tilde\Gamma,g)^2
\eea
where
\bea\label{e02}
R(\tGamma,g)=g^{\mu\nu}\,R_{\mu\nu}(\tGamma),\qquad
R_{\mu\nu}(\tGamma)=\partial_\lambda\tilde\Gamma^\lambda_{\mu\nu}-
\partial_\mu \tGamma^\lambda_{\lambda\nu}
+\tGamma_{\rho\lambda}^\lambda \tGamma^\rho_{\mu\nu}-\tGamma^\lambda_{\rho\mu}
\tGamma^\rho_{\nu\lambda}.
\eea
%\medskip\noindent
This is the Palatini version of  action (\ref{L1}) in Weyl geometry, but now $\tilde\Gamma$ is unknown.
$R_{\mu\nu}(\tGamma)$ is the metric-independent Ricci tensor in the Palatini formalism.
Since $\tGamma$ is  independent  of the $g_{\mu\nu}$, 
$\tGamma$ and $R_{\mu\nu}(\tGamma)$ are invariant under transformation (\ref{WS}).
Therefore,  $R(\tilde\Gamma, g)$ transforms like $g^{\mu\nu}$, so under (\ref{WS}):
\bea
\hat R(\tGamma,\hat g)=\frac{1}{\Sigma} R(\tGamma,g).
\eea
%\medskip
As a result, $\cL_2$ is  invariant under (\ref{WS}) and this is the simplest
Palatini case that has local scale symmetry (\ref{WS}).
$\cL_2$ can be linearised as in the Weyl case (eq.(\ref{oo})): replace
$R(\tilde\Gamma,g)^2 \ra -2 \phi^2  R(\tilde\Gamma,g)-\phi^4$ to obtain a classically equivalent action
to (\ref{L2}).
For this  $\cL_2$ one  writes and solves the equation of motion for $\tilde \Gamma$.
The  solution is (see \cite{Palatini1}, Section~2):
\medskip
\bea\label{on}
\tGamma_{\mu\nu}^\alpha=\Gamma_{\mu\nu}^\alpha(g)
+
(1/2)\,
\big(\delta_\nu^\alpha \,u_\mu+\delta_\mu^\alpha u_\nu
-g^{\alpha\lambda} g_{\mu\nu} u_\lambda\big),\quad u_\mu\equiv\partial_\mu \ln\phi^2,
\eea

\medskip\noindent
with $\Gamma$ the  Levi-Civita connection. 
With this $\tilde\Gamma$, one  computes the scalar curvature
 \smallskip
\bea\label{rel}
R(\tGamma,g)=R(g)- 3 \nabla_\mu u^\mu-\frac{3}{2} \, g^{\mu\nu} u_\mu \, u_\nu.
\eea

\smallskip\noindent
 $R(g)$ is the scalar curvature for $g_{\mu\nu}$ while $\nabla$ is defined by the 
 Levi-Civita connection ($\Gamma$). Using the last equation back in $\cL_2$, one finds
\medskip
\bea\label{rt}
\cL_2=\sqrt{g}\,  \frac{1}{\xi^2}\,
\Big\{-\frac12 \,\Big[\, \frac{1}{6}\, \phi^2 R(g) + (\partial_\mu\phi)^2\Big]
-\frac{1}{4!} \,\phi^4\Big\}.
\eea

\medskip\noindent
This is the onshell version of (\ref{L2}) and contains a dynamical $\phi$.
This is because the metric part of the
Palatini quadratic gravity\footnote{due to the Levi-Civita contribution to $\tilde \Gamma$,
  eq.(\ref{on}).} makes the action a four-derivative theory: according to
(\ref{rel}) onshell $\tilde R(\tGamma,g)^2$ contains $R(g)^2$.
Action (\ref{rt}) is similar to that  seen  in
eqs.(\ref{RM}) and (\ref{RMW}) and its associated current vanishes again.
Fixing  the gauge of the symmetry which essentially means setting $\phi\ra \langle\phi\rangle$ then
%\medskip
\bea
\cL_2=\sqrt{\hat g}\,\, \Big\{ \frac{-1}{2} M_p^2 \hat R(\hat g) -\frac{3}{2\xi^2}\,M_p^4\Big\}
\eea

\medskip\noindent
And since  $\phi$ is fixed to a constant, $\tilde\Gamma=\Gamma$, so the theory
is {\it metric}, see e.g. discussion in \cite{Palatini1}. This is similar to the
Weyl integrable (metric) case discussed earlier.

\bigskip
\noindent
{\bf $\bullet$ Non-metric case:\,\,\,}

\medskip\noindent
The situation changes dramatically  if the theory has a gauged scale invariance.
Consider
\smallskip
\bea\label{L2p}
\cL_2'=\sqrt{g}\,\Big\{\frac{1}{4!\,\xi^2} R(\tilde\Gamma,g)^2
- \frac{1}{4 \q^2}\,R_{[\mu\nu]}(\tilde\Gamma) \,R^{[\mu\nu]}(\tilde\Gamma)\Big\}
\eea

\smallskip\noindent
where $R_{[\mu\nu]}\equiv (1/2)\,(R_{\mu\nu}- R_{\nu\mu})$ with $R_{\mu\nu}$ as in eq.(\ref{e02}).
Additional scale invariant operators of dimension d=4 can be present.
One can check that the two terms in the action above are invariant under (\ref{WS}).
Next, define
%\medskip
\bea\label{c2}
F_{\mu\nu}(\tilde\Gamma)=\tilde\nabla_\mu v_\nu-\tilde\nabla_\nu v_\mu,
\quad
\text{where}
\quad v_\mu\equiv (1/2) (\tilde\Gamma_\mu -\Gamma_\mu),
\quad
(\tilde \Gamma_\mu\equiv \tilde\Gamma_{\mu\rho}^\rho,
\,\,\Gamma_\mu\equiv \Gamma_{\mu\rho}^\rho).
\eea
%
%\medskip\noindent
so $F_{\mu\nu}$ is a function of $\tilde\Gamma$.
Since $\tGamma$ is assumed symmetric in the lower indices,
then $F_{\mu\nu}=\partial_\mu v_\nu -\partial_\nu v_\mu
=\partial_\mu \tilde\Gamma_\nu -\partial_\nu \tilde\Gamma_\mu
=-R_{[\mu\nu]}$. Hence, the last term in (\ref{L2p}) acts as a 
kinetic term for $v_\mu$.
Under (\ref{WS}), $v_\mu$ transforms like $\w_\mu$ of the Weyl case,  while
$F_{\mu\nu}$ is invariant. Therefore, action (\ref{L2p}) has a bigger symmetry: it is
Weyl gauge invariant, being invariant under eqs.(\ref{WS}) and (\ref{WGS}) with $\w_\mu\ra v_\mu$
i.e. $\hat v_\mu=v_\mu-(1/\alpha)\,\partial_\mu\ln\Sigma$.
We denoted by $v_\mu$ the  Weyl gauge field in the Palatini case,
playing the role of $\w_\mu$.
With this, eq.(\ref{L2p}) 
is a Palatini version of  the  Weyl action in eq.(\ref{inA})
but now $\tilde\Gamma$ is {\it unknown} - it
will be determined by its equations of motion.

From (\ref{L2p}) one proceeds as in the Weyl case to ``linearise'' the $R(\tGamma, g)^2$
term in (\ref{L2p}) with the aid of an auxiliary
scalar $\phi$, so replace $R(\tilde\Gamma,g)\ra -2\phi^2\,R(\tilde\Gamma,g)-\phi^4$.
From the resulting, equivalent Lagrangian one can then write the equations of motion
of the connection $\tilde\Gamma_{\alpha\beta}^\rho$ which can be solved.
The solution is a function of $\phi$ and is used to
evaluate $R_{\mu\nu}(\tilde\Gamma)$ and then $R(\tilde\Gamma,g)$.
Using this result  back in action (\ref{L2p}) one finally finds  \cite{Palatini1} (eq.29):
%\medskip
\bea\label{opop}
\cL_2^\prime=
\sqrt{g}\,\Big\{
-\frac{1}{2\,\xi^2} \Big[\frac{1}{6}\,\phi^2\,R
+ (\partial_\mu\phi)^2\Big]
- \frac{1}{4!\,\xi^2} \phi^4
+ \frac{\q^2}{2\xi^2} \,\phi^2\,\Big[v_\mu-\frac{1}{\q^2}\partial_\mu\ln\phi^2\Big]^2
- \frac{1}{4 } F_{\mu\nu}^2 \Big\},
\eea

\medskip\noindent
This result is similar to that in Weyl case eq.(\ref{alt2}), with $v_\mu\ra\omega_\mu$.
Eq.(\ref{opop}) is the ``onshell'' Lagrangian, that is using the solution of $\tilde \Gamma$.
$\cL_2^\prime$ is invariant under combined eqs.(\ref{WS}), (\ref{WGS})
(with $\w_\mu\ra v_\mu$). A conserved current exists similar to that in
    Weyl case eq.(\ref{current}) and Appendix~A.

To fix the gauge, the same discussion as in the Weyl case applies.
At the level of the Lagrangian
this ``gauge fixing'' may formally be  implemented  by setting $\phi$ to a constant vev;
this then brings us to the Einstein-Proca Lagrangian
%\medskip
\bea\label{yyy}
\cL_2^\prime=\sqrt{g}\,\Big\{-\frac12 M_p^2\,R + 3 \q^2\, M_p^2 v_\mu\,v^\mu
-\frac14\,F_{\mu\nu}^2
-\Lambda\,M_p^2
% -\frac32 \xi^2 M_p^4
\Big\},
\qquad \Lambda\equiv\frac14\langle\phi\rangle^2,
\quad M_p^2\equiv \frac{\langle \phi\rangle^2}{6 \,\xi^2}.
\eea

\medskip\noindent
There is again a Stueckelberg mechanism, similar to the Weyl case:
$v_\mu$ becomes massive after ``absorbing'' the dynamical $\phi$ which disappears from
the spectrum of (\ref{yyy}) and the
number of degrees of freedom is conserved in going from (\ref{opop}) to (\ref{yyy}).
The Einstein-Proca action of $v_\mu$ is thus found. We see that
the Stueckelberg breaking of a gauged scale symmetry is  valid in a
Palatini quadratic gravity model, too. This  mass mechanism may be common in gravity theories
  where the connection is a dynamical variable \cite{ST2,ST3}.

There are however two differences from the Weyl case:
Firstly,  there are additional quadratic operators \cite{Mar} with gauged scale invariance
that were not included in this analysis and that can affect the overall
result.
Secondly,  the vectorial non-metricity obtained in the Palatini case for action (\ref{L2p}),
shown in the Appendix
eq.(\ref{pg}), is different from that in the  Weyl case eq.(\ref{nm}). This explains
a different numerical coefficient in (\ref{opop}) versus (\ref{alt2}).

The Weyl and Palatini cases above  show that non-metricity that follows from 
gauging the  scale symmetry is accompanied by a
Stueckelberg breaking of this symmetry to Einstein-Proca action\footnote{
  For a related scale invariant de Sitter gauge theory see \cite{Koivisto}.}.
They have further advantages compared  to their metric versions: a)
they have  a non-trivial conserved current $J^\mu$ 
and  b) there are no ghost fields in  actions (\ref{EP}), (\ref{yyy}).

\section{Phenomenology}\label{pheno}

\subsection{Mass scales from non-metric geometry}

We saw that Einstein gravity is a  spontaneously broken phase of
the original Weyl quadratic gravity or  its Palatini version, with Weyl gauge symmetry.
We have 3 mass scales: $M_p$, $m_\w$,  $\Lambda$
that are all
proportional to the vev of the Stueckelberg field   $\langle\phi\rangle$ eaten by $\omega_\mu$ ($v_\mu$),
eqs.(\ref{massesW}), (\ref{yyy}). Their exact values are fixed by three
parameters: $\langle\phi\rangle$, $\xi$ and $\alpha$.
Here $\phi$ is introduced by a geometric term in the action ($\tilde R^2$), so all masses
have a {\it non-metric geometry} origin!   $\w_\mu$ ($v_\mu$) is also part of the underlying non-metric
geometry, too. There  is a difference from the  Higgs mechanism, since there is no $\phi$ present
in the final action.
The  cosmological constant is positive (and non-vanishing), due to a
$\phi^4$ term also induced by $\tilde R^2$. This suggests  a UV - IR  physics connection 
due to the common  origin ($\propto \langle\phi\rangle$) of the scales  $\Lambda$ and $M_p$. 

As in all theories with scale symmetry one can only predict ratios of scales
in terms of dimensionless couplings of the theory ($\xi$, $\alpha$).
One can  obtain the correct ratios
%\medskip
\bea
\Lambda/M_p^2\sim \xi^2,
\qquad
m_\w^2/M_p^2\sim \q^2
\eea
%
%\medskip\noindent
for suitable (perturbative) values of $\alpha$ and $\xi$.
If the Planck scale $M_p$ is fixed to its value, we see that $\Lambda$  is small because gravity
is ultraweak (coupling $\xi\ll 1$).

We also see that issues {\bf a), b), c), d)} encountered in the  (pseudo-)Riemannian
case  with Weyl symmetry of  Section~\ref{metric} are now nicely solved.  To detail:
{\bf a)} There is no ghost degree of freedom in the final spectrum, since $\phi$ is eaten by $\w_\mu$.
Also, the Einstein gravity is recovered and the
sign of $\Lambda$ is predicted  positive, due to the $\sqrt{g}\tilde R^2$ term;
{\bf b)} The current associated to Weyl gauge symmetry
is non-trivial ($J^\mu\not=0$) and is related to the existence of a dynamical
$\w_\mu$ i.e. to non-metricity;
{\bf c)} The field $\phi$ was not added by hand, but was
``extracted'' from the  $\sqrt{g}\tilde R^2$ term, hence it has an origin in  (non-metric) geometry
 and the same is true about $\Lambda$, $M_p$ and $m_\omega$ that $\langle\phi\rangle$ generated;
{\bf d)} Finally, the Weyl or Palatini connections are Weyl gauge invariant,
hence both  the action and its underlying geometry have this symmetry.

\subsection{Higgs from non-metric geometry}

The discussion so far was in the absence of matter, hence it was  about ``geometry''.
The next step is to see the effect  of non-metricity in the presence of the SM. Consider then
embedding  the SM in Weyl conformal geometry -  this is indeed possible,
as shown in \cite{SMW}.
We refer the reader to this work for the technical details how this is done.
This embedding is truly minimal and natural and
does not require any additional degrees of freedom beyond those of the SM and of
Weyl geometry ($\phi$, $\w_\mu$, $g_{\mu\nu}$).
This is possible because the SM with a vanishing Higgs mass parameter is scale
invariant.
In fact, one can
easily notice that the Lagrangian of the SM gauge bosons and fermions
in the Weyl conformal geometry has a form  identical to that in the (pseudo-)Riemannian geometry
and has a gauged scale symmetry.
Hence,  the SM gauge  bosons and  fermions do
not have any direct couplings to the Weyl gauge boson  \cite{Kugo} (with one special exception
for the SM fermions discussed in Section 2.3 of \cite{SMW}).

However, the  SM  Higgs sector  is modified by the Weyl gauge symmetry \cite{SMW}.
First, there is a non-minimal coupling of the Higgs to Weyl geometry
$\sqrt{g}\,H^\dagger H \tilde R$ which is Weyl gauge invariant;
here $H$ is the $SU(2)_L$ Higgs doublet.
The Higgs kinetic term is also modified: the SM  covariant derivative
$D_\mu H$ is ``upgraded'' to also become Weyl-covariant; hence
the  derivative $D_\mu H$ is replaced to include the Weyl
gauge boson of scale invariance:
%\medskip
\bea
D_\mu H \ra (D_\mu -\q/2\,\w_\mu) H,
\eea
so that this derivative transforms like the Higgs  under
transformations (\ref{WS}), (\ref{WGS}).
Hence, the Weyl gauge invariant Higgs kinetic term becomes
\bea
\sqrt{g}\,g^{\mu\nu}\,  \big[ D_\mu -\q/2\,\w_\mu)\,H\big]^\dagger
(D_\nu -\q/2\,\w_\nu)\,H.
\eea
Further,  the neutral Higgs boson  mixes  with the field $\phi$ that
``linearised'' $\tilde R^2$; their ``radial direction'' combination is
now the new  Stueckelberg field  eaten by $\w_\mu$ (as  shown earlier),
while the ``angular direction''  field becomes  the SM neutral Higgs, hereafter called $\sigma$.

In the canonical Lagrangian, the  coupling of $\w_\mu$ to $\sigma$ is
(see  eqs.(32), (38) in \cite{SMW}):
%\medskip
\bea
\cL_H=\frac18\,\sqrt{g}\, \q^2\,\w_\mu \w^\mu\,\sigma^2+\cO(\sigma^2/M_p^2).
\eea
% \medskip\noindent
This coupling  comes from the Higgs kinetic term mentioned earlier\footnote{
  The coupling $\sqrt{g}\, H^\dagger H \tilde R$
 impacts  on the Higgs potential and the mixing with initial $\phi$,
  see \cite{SMW} Section~2.5.}.
This  is the only direct coupling of the SM to the Weyl gauge boson. It has interesting
consequences.
In the early Universe, assuming there was no Higgs boson,
this coupling can generate the Higgs  from the
Weyl vector boson fusion
%\medskip
\bea
\w_\mu+\w_\mu\ra \sigma+\sigma.
\eea
%\medskip\noindent
Since  $\w_\mu$ is part of the non-metric geometry (connection),  the
Higgs boson itself  can have a non-metric,  geometric origin! 
And since the Higgs generates the masses of the SM states  while the Stueckelberg field
(``extracted'' from the $\tilde R^2$ term) generated $M_p$, $\Lambda$
and  $m_{\w}$, one concludes that
$\w_\mu$ and non-metric geometry  are the origin of  all the masses
of the theory. This happens without additional degrees
of freedom beyond the SM or Weyl geometry!

Interestingly, the Weyl  boson fusion can have an additional effect at a cosmological level,
of mitigating any anisotropy that the Weyl vector would otherwise  bring.
This deserves careful study.  A similar coupling and generation of the Higgs  via Weyl boson fusion
exists when considering the Higgs sector in Palatini  gravity
with Weyl gauge symmetry \cite{Palatini1} (eq.45).

The generation of the  Higgs field alone, from  non-metric geometry (Weyl or Palatini), via
$\w_\mu$-$\w_\mu$ fusion in the early Universe is interesting\footnote{Matter
  generation from metric geometry is however familiar: e.g. MSSM states as zero modes
  of string compactifications \cite{Z6},  Kaluza-Klein modes of SM states
  in field theory orbifolds, gravitational particle production during inflation, etc.}.
This process appears as a possible microscopic realisation of
 ``matter creation from geometry'' discussed in a phenomenological
macroscopic description in  the thermodynamics of open systems applied to cosmology
\cite{P1,P2}. In such approach the  creation of matter occurs as a process
corresponding to transfer of energy from the gravitational field(s)
(in our case $\w_\mu$) or  space-time curvature ($\tilde R^2$ that depends on $\w_\mu$)
to  the matter created (in our case Higgs).
The second law of thermodynamics allows space-time geometry transform into matter
but the inverse transformation is forbidden. Therefore the process
of matter creation from the underlying geometry is irreversible.
However, this result is
 not valid in general but only  if the specific entropy per particle ($s$)
 is $\dot s \leq 0$ (otherwise matter destruction takes place) \cite{P4}. 
If so, it would be interesting  to  study  how irreversibility could emerge from
the microscopic picture provided by our  SM Lagrangian in Weyl
geometry. Notice however that what from the Weyl geometry viewpoint  
looks like matter creation from (Weyl) geometry ($\w_\mu$), from a Riemannian picture
obtained after the symmetry breaking,
$\w_\mu$ looks just like another  field of the theory that interacts with the Higgs!
Similar considerations apply in the Palatini case.

\subsection{Non-metricity and mass hierarchy}

With the Higgs and Planck scale related to the underlying non-metric geometry,
their hierarchy may be be  related to this, too.
The scale of non-metricity is given by the mass of the gauge boson of scale invariance,
$m_\w\sim \q M_p$; below this scale metricity is restored. In general, one would expect that
this scale be close to the Planck scale.  But it must be mentioned that the
current lower bound on the non-metricity  scale is very low, of few TeV only \cite{Latorre}. 
Theoretically, this is realised by tuning the coupling  $\q\ll 1$.
Such small   $\q$ is actually natural, because  it is one of the
gravity couplings of the theory ($\xi, \q$).  For detailed numerical estimates of
the couplings $\xi$, $\q$ and Higgs non-minimal coupling, due to constraints from
EW precision data and inflation, see \cite{SMW}.

If the non-metricity scale $m_\w$ is low, in the TeV
region, then the Higgs mass is natural, as already noticed in \cite{SMW}.
To see this note that quantum corrections $\delta m_\sigma^2$
to the Higgs mass are quadratic in the scale of ``new physics'' (in our case $m_\w$), so
\bea
\delta m_\sigma^2\sim m_\w^2.
\eea
%\medskip\noindent
Above the mass of $\w_\mu$, the gauged scale symmetry is restored together with its UV  protection
for the Higgs mass; indeed, this is so since no mass counterterm is then
allowed and quantum corrections above $m_\w$
could at most be of logarithmic type. This indicates  a solution to the hierarchy problem that
is technically natural, based on a gauged scale symmetry\footnote{It would also be
  interesting to construct  a supersymmetric version of Weyl quadratic action (\ref{inA}) - to our
  knowledge there is no such version in the current literature.}.

\subsection{Non-metricity in solid state physics}

We saw that there is a link of non-metric geometry to mass generation
which is an essentially geometric mechanism, valid even in the absence of  matter fields;
only the degrees of freedom of the Weyl geometry,  like the metric, connection
and curvature-squared terms were involved. The same applied to  the Palatini case.
One question is whether there is a more intuitive, physical interpretation of non-metricity
and its link to mass generation in the condensed matter physics, as it was
the case for the Higgs mechanism.

Non-metricity is common in solid state physics where it is associated with
some crystalline defects.
For our discussion on  non-metricity in solid state physics we follow \cite{cond1,cond3,cond4}.
For this one needs the notion of ``material space'' of a crystalline solid.
This is a natural configuration of a body where it is described only in terms of the
intrinsic structure of constituting matter. The  material space
is the configuration  found  by relaxing the solid of all internal
and external stresses. If the crystalline structure has no defects, the corresponding
geometry of this material space is Euclidean. If there is a (continuous) distribution of
defects, that destroy the crystalline order, the associated geometry is non-Euclidean.

For a  3D crystalline structure we have defects of dimension
d=0, also known as point defects or metric anomalies, which
 are destroying the crystalline order;  they are  modifying the local
notion of length,  usually associated with this order. These defects  can be vacancies
(missing atoms), interstitials (extra atoms of same kind), substitutionals (extra atoms
of different kind). Further, there are d=1 defects such as dislocations and disclinations;
d=2 defects (phase boundary, domain walls, etc) or d=3 defects (inhomogeneities).
With these defects distributed continuously, they give rise  to effective fields of defect densities.

The material space is then  described geometrically by an affine connection
of a non-Riemannian space that has  non-vanishing curvature,  non-metricity and torsion
\cite{cond1,cond3}. Non-metricity, which we know by definition  it modifies the local notion of length
must therefore  be  related to a density of the  d=0 defects.
Then the relation of mass generation   to non-metricity that we found
is somewhat expected, given that mass terms in the action break  the
local Weyl symmetry of the action, much like d=0 defects destroy the local order
of the crystalline.  At low energy we saw that metricity is recovered in our case,
just like the local d=0 crystalline defects are not observable from
large distances relative to the  lattice size.
This view  gives an intuitive picture to non-metricity and its relation
to mass generation.

Further, the  curvature tensor is associated with a density of disclinations.
In the absence of torsion (as in our Weyl and Palatini cases) the material
connection is similar to Weyl connection and the geometry is then Weylian.
If present, torsion is associated with dislocations and the geometry
is modified.  The material connection can then be expressed in
terms of the torsion, non-metricity and  metric. Note there is a
physical distinction of these concepts, something less obvious in general
gravity theories \cite{GTG}. This concludes our brief description of
curvature, torsion and non-metricity from a solid state physics perspective.

\subsection{The multiple roles of the  Stueckelberg field ($\phi$)}

It is worth noting the multiple role of the scalar mode
$\phi$ of $\tilde R^2$  in Weyl and Palatini cases:

\noindent
{\bf 1)} It acts as a Stueckelberg field eaten by $\w_\mu$,

\noindent
{\bf 2)} Its vev generates $M_p$, $\Lambda$, $m_\w$, and it is 
playing the role of the dilaton - indeed, note that $\ln\phi$ has a shift symmetry
under  transformation (\ref{WS}).

\noindent
{\bf 3)} When   SM is embedded in Weyl geometry, it mixes with the Higgs leading to a
Stueckelberg-Higgs potential which for small Higgs field
values recovers the SM potential~\cite{SMW}.

\noindent
{\bf 4)}  The contribution of $\phi$ to this potential 
drives inflation in such $\tilde R^2$ models 
\cite{Palatini1,Ferreira,Winflation,SMW}. This  explains why the inflation prediction
for the tensor-to-scalar ratio ($r\sim 10^{-3}$) in Weyl case is
similar to that in the Starobinsky model \cite{Sta}  ($r$ is larger  in the  Palatini case due to different
coefficients in the scalar potential caused by different vectorial non-metricity) \cite{WvsP}.

\subsection{Renormalizability}

In a most general case, the action  in the Weyl theory (\ref{inA})
can include (up to a topological term) only one  additional operator  that
also has a gauged scale invariance.
This is $\tilde C_{\mu\nu\rho\sigma}^2$, where $\tilde C_{\mu\nu\rho\sigma}$
is the Weyl tensor in Weyl geometry.
This is  related to the usual Riemannian Weyl tensor
$C_{\mu\nu\rho\sigma}$ via  $\tilde C_{\mu\nu\rho\sigma}^2= C_{\mu\nu\rho\sigma}^2 + (3/2) \alpha^2 F_{\mu\nu}^2$
 where $F_{\mu\nu}^2$ is  the kinetic term of $\w_\mu$.
The operator
$\tilde C_{\mu\nu\rho\sigma}^2$ is essentially spectator under
the mechanism of symmetry breaking presented earlier and does not affect the results shown.
In a quantum analysis, it might be generated as a loop counterterm. 
The Riemannian version of this term ($C_{\mu\nu\rho\sigma}^2$)
was extensively analysed \cite{Mannheim},
while the extra $F_{\mu\nu}^2$ contribution only  brings a redefinition of coupling $\alpha$.

Given the gauged scale invariance of the action, there are no  operators of dimension larger than four
that can be present in the action, since there is no scale  to suppress them.
In a  Riemannian notation, the overall Lagrangian thus
involves only $(R-3\alpha\nabla_\mu\w^\mu-3/2\, \alpha^2\,\w_\mu \w^\mu)^2$ coming from $\tilde R^2$
that is linearised with the aid of $\phi$, then 
$F_{\mu\nu}^2,$ and   $C_{\mu\nu\rho\sigma}^2$. 
The Weyl vector is massive and the symmetry is {\it  anomaly-free}   \cite{SMW,anomaly}, with $\omega_\mu$
of mass acquired via spontaneous breaking  which  cannot affect the renormalizability of the theory. 
Further, it is known that the usual quadratic gravity in the
(pseudo)-Riemannian case is renormalizable \cite{Stelle}.
For the Weyl theory,  based on the symmetry of the action forbidding higher dimensional
counterterms, the analysis of \cite{Stelle} and power-counting arguments, one
expects  this theory be renormalizable (but not unitary, due to spin-2 ghost of $C_{\mu\nu\rho\sigma}^2$).

In the Palatini approach a similar
discussion is difficult: there are many additional Weyl gauge invariant operators that
can be present in  the action \cite{Palatini1,Mar}, then solving  analytically
the equations of motion for the connection and  finding the non-metricity is very  difficult.

\subsection{Non-metricity: Weyl vs Palatini}\label{38}

There has been a long held view since Einstein's critique  \cite{Weyl1} that non-metricity
(i.e. $\tilde\nabla_\mu g_{\alpha\beta}\!\not=\!0$) makes
a theory unphysical\footnote{This was the original critique of Einstein
  to Weyl's failed theory of gravity ``plus'' electromagnetism.}. Firstly, since
$\tilde\nabla_\mu g_{\alpha\beta}\not=0$ \footnote{For a discussion of metric versus non-metric theories
  see \cite{nonm}.} it is usually stated that
under the parallel transport of a vector,  such vector
changes not only the direction as in the Riemannian geometry, but also its norm.
Hence, the norm of  a vector or the clock rate are usually path dependent
(there are actually exceptions to this result, as reviewed in  Appendix~B
under the assumption of a massless $\omega_\mu$ and  Weyl gauge symmetry  present).
In any case, the usual critique relates to the physical
consequence of changing the norm of a vector or the clock rate;   in an experiment
one would conclude  that the distance between the spectral lines of two identical atoms
of different  path history will then  differ, in contrast to the  experience
(the so-called second clock effect) \cite{Weyl1}.
  In the light of our result
that $\omega_\mu$ is actually {\it massive}, this claim  must be reviewed.

In our view the above critique (reviewed recently in \cite{Quiros})
is implicitly assuming  a formalism which
actually breaks the Weyl gauge symmetry of  the action; this
is seen when setting the momenta on the
mass shell $p^2=m^2$ as done in  \cite{Quiros}.
%%% 
This  means that a mass term is actually present in the action  but
that means  we are actually in a {\it broken phase},
or we  already explained that  in a broken phase
the theory is  {\it metric}, hence the formalism and critique cannot apply.

More generally, in the  symmetric phase of the theory i.e. without masses or
other dimensionful couplings present in the action, it is difficult
to explain how the above experiment could actually be physically
realised. Indeed,  if there is no mass scale in the theory,  one cannot define 
a clock rate in that theory! So the critique cannot apply.  Secondly, 
comparing a gauge theory (of scale invariance in our case) to the experiment first requires a 
``gauge fixing'' of this symmetry.
From  the equations of motion of the Weyl field \cite{SMW}
and  Weyl current conservation $\nabla_\mu J^\mu=0$
\cite{Ghilen1,SMW}, the  gauge fixing  condition ($\nabla_\mu \w^\mu\!=\!0$)
follows after $\phi$ acquires a non-zero vev, which  in turn breaks this symmetry.
Hence we are  back to the  broken phase of the theory which is  {\it metric}, 
$\omega_\mu$ is massive, decouples and  the original critique of Einstein cannot apply!
This line of reasoning   implies  that in Weyl's original theory
the second clock effect is not there or,
more exactly,  it is  suppressed by a (large) mass $m_\w$ as first shown for this theory
in \cite{Ghilen1}. Hence,  our results are not affected by this critique.

To summarise,  we know that the gauged scale  symmetry is broken, 
both in  the Weyl quadratic gravity and in the Palatini  case.
When this symmetry is broken,  the massive $\w_\mu$  decouples (at some high scale)
from the Lagrangian and its underlying geometry:
the connection becomes Levi-Civita and the theory is then metric;
any non-metricity effects and implications  mentioned above (if present) are  then
suppressed by $m_\w$. As long as $m_\w$ is large enough, such effect
can be ignored. As mentioned,  the current lower bound on non-metricity ($m_\omega$)
derived from its effect on $e^+ \, e^- \ra e^+ \,e^-$ scattering,  is actually very low, 1 TeV
\cite{Latorre}!

 Finally, we would like to comment on the different vectorial
 non-metricity of Weyl versus the Palatini case,  compare eq.(\ref{wg}) to (\ref{pg}).
 This  has  implications for the parallel transport of the vectors.
 As shown in  Appendix~B, in Weyl geometry the ratio of the norms
 of two vectors $u^\mu, t^\mu$ of equal, non-zero, arbitrary Weyl charge
 is invariant under their parallel transport along the same curve
\bea\label{qqqq}
d\,\frac{\vert u\vert^2}{\vert t\vert^2}=0.
\eea
So the {\it relative} length is  invariant and this is consistent with physics
being independent of the units of length. Actually, in theories based on (non-metric)  Weyl geometry
even the  norm itself  of a vector
$u^\mu$ is invariant under a parallel transport on any curve,
if its tangent space version $u^a\!=\!e^a_\mu \, u^\mu$
is invariant (has zero charge), see eq.(\ref{norm}) in Appendix~B; this result  applies even
though  the theory has non-metricity $\tilde\nabla_\mu g_{\alpha\beta}\not=0$.

In the Palatini case, however, neither the norm of a vector nor the ratio of the norms of two vectors
are invariant, as seen from eq.(\ref{ratioP}) in  Appendix~B
 \bea\label{ff}
 d\,\frac{\vert u\vert^2}{\vert t\vert^2}\not=0.
 \eea
 Compared to (\ref{qqqq}), one may conclude on physical grounds that the Weyl case is
 the only acceptable.  We note, however, that  the Palatini case is affected
 by additional operators not included in our analysis that can change (\ref{ff}).
 It may even be possible that a most general Palatini quadratic gravity with gauged scale invariance
 may yield onshell (after solving the equations of motion for the connection)
 a Weylian non-metricity and connection.
This would give an interesting offshell realisation of  Weyl quadratic gravity.

\section{Conclusions}\label{conc}

We discussed phenomenological aspects of  non-metricity  in theories
beyond the SM and Einstein gravity that have a {\it gauged} scale symmetry.
One  argument in favour of this symmetry is the gauge principle:
similarly to the SM as a gauge theory, we seek a gauge theory of scale invariance 
that recovers Einstein gravity in its broken phase.

What is the 4D underlying geometry of such theories?
One can consider theories based on the Weyl  geometry which has
a {\it gauged} scale symmetry built in  i.e. the Weyl connection has this symmetry. 
A second option is to consider the Palatini approach to  gravity in which the (offshell)
connection of the underlying geometry also  has this symmetry, being independent of the metric
and its Weyl transformation.
The consequence  is that the underlying geometry of these
theories is  non-metric i.e.  $\tilde\nabla_\lambda g_{\mu\nu}\not=0$; 
 in other words,  non-metricity   is a result of gauging the scale symmetry ($\omega_\mu$ dynamical).
 This situation is different from  theories in which $\omega_\mu$ is not dynamical,
 based on the (metric) pseudo-Riemannian geometry with a local scale  symmetry (no $\omega_\mu$ present)
under which its  Levi-Civita connection is {\it not} invariant - in such case the geometry
does not share the space-time symmetry of the action - which raises concerns
about  their consistency. % 

Rather than  being a problem (as it was thought in the past),
 non-metricity of Weyl or Palatini cases  plays a crucial
role in mass generation in gauge theories of scale invariance:
it brings  mass generation  via a 
Stueckelberg breaking of this symmetry in Weyl or Palatini quadratic gravity
{\it in the absence of matter}. There are additional advantages of non-metricity.
Firstly, there is a non-trivial, conserved current associated with
the Weyl gauge symmetry and secondly, there are no ghost degrees of freedom in the action with
this symmetry. This is unlike (metric or integrable) theories with local scale symmetry
only (no dynamical $\omega_\mu$) where  Weyl current is trivial, as shown by
Jackiw and Pi \cite{J1,J2} and a  ghost  is present.

Our results show that in  the absence of matter,
{\it all mass scales have a  geometric origin:}
the Planck mass, the cosmological constant $\Lambda$ and the mass of $\w_\mu$ 
are  proportional to $\langle\phi\rangle$ which
is  the spin-zero mode propagated by the geometric  $\tilde R^2$  term.
Unlike in local scale invariant theories (no  $\w_\mu$ present)
based on the  (metric) pseudo-Riemannian geometry, in Weyl and Palatini 
 cases  we studied   {\it no} scalar fields were added ``ad-hoc'' or needed
to generate  these mass scales.
A hierarchy of  scales ($\Lambda$, $M_p$, $m_\omega$)  is related
to the smallness of the dimensionless gravitational
couplings $\xi$, $\alpha$ at a classical level: with $M_p$ {fixed}, the cosmological constant 
is small because gravity is weak ($\xi\ll 1$).  There is a UV - IR physics connection
associated with $M_p$ and $\Lambda$, respectively, since these  scales have a common origin
($\propto\langle\phi\rangle$). Finally, $\Lambda>0$ because it is due to the
$\phi^4$ term  induced again by geometric $\tilde R^2$.

Metricity is recovered below $m_\w$ after the massive Weyl gauge boson of
scale symmetry decouples from the spectrum. In this decoupling limit the
connection becomes Riemannian (Levi-Civita) and the geometry is metric.
The scale where this happens ($m_\w$) is naively expected to be high
($\propto\alpha M_p$), but current bounds on the non-metricity scale are actually very low (TeV scale). 
A low value of $m_\w$  can be realised for a small coupling $\alpha\ll 1$.
These results, obtained in the absence of matter,
also  apply to the Palatini case.

The above  picture remains valid if matter is present, when the SM (with a massless higgs)
is embedded in Weyl geometry. This is a natural embedding, without new degrees of freedom
beyond the SM and Weyl geometry.  
Of the SM spectrum  only the Higgs field ($\sigma$) has a direct coupling to
$\w_\mu$ of the form $\w_\mu \w^\mu \sigma^2$.
This leads to the interesting possibility that the Higgs  be generated
by Weyl vector fusion $\w_\mu+\w_\mu \ra \sigma+\sigma$ in the early Universe.
Since $\w_\mu$ has  geometric
origin, this  means that the Higgs itself has an origin in  Weyl's non-metric  geometry, too.
Therefore, not only the scales of quadratic gravity are  of geometric origin but
this extends, in a sense, to all SM  masses generated by the Higgs.
This shows  that Weyl geometry is more fundamental and it provides a
UV completion of the (pseudo)Riemannian geometry; correspondingly,  the associated
Weyl quadratic gravity provides a gauge theory embedding of Einstein gravity.
These results also apply  to the Palatini case; however,  in this case
there are unknown corrections from additional operators not included in our study.

Non-metricity is  common in solid state physics where it is associated with
crystalline structure defects of dimension d=0 (point defects or metric anomalies).
They  destroy the crystalline order and modify the local notion of length
associated with this order. Then the relation of mass generation to  non-metricity
that we found is  expected, given that mass terms in the action break the local
(Weyl) symmetry much like point defects destroy the local order/size of the lattice. At low energy
we saw that metricity is recovered, much like the local crystalline defects are not observable
from large distances relative to the lattice size. This gives an intuitive picture of non-metricity.

Can these ideas about non-metricity be tested experimentally? 
One possibility is to analyse a possible imprint on the  gravitational
waves due to the Weyl gauge boson of scale symmetry.  The second
possibility is in Higgs physics,   assuming a light  $\w_\mu$ near its lower bound;
in this case the term  $\w_\mu \w^\mu \sigma^2$, relating Higgs physics to non-metricity,
brings corrections to the Higgs couplings (e.g. quantum corrections to the quartic coupling).
In this way one may  set lower bounds on $m_\w$ which is the scale of ``new physics''
in this case. A third possibility is via the Stueckelberg-Higgs inflation,
which predicts a  low ($\sim 10^{-3}$) tensor-to-scalar ratio ($r$) value,
testable in the near future experiments.
Work to explore  these interesting possibilities is in progress.

%\bigskip
\bigskip
\begin{center}
  ------------------------------------
\end{center}

%\vspace{0.5cm}

%\newpage
 \section*{Appendix}

\def\theequation{A-\arabic{equation}}
\def\thesubsection{A}
\setcounter{equation}{0}
\def\thefigure{A-\arabic{figure}}
\def\thelabel{A}

 \subsection*{A. Weyl gauge invariant theories  and conserved current}

$\bullet$  For an arbitrary Weyl gauge invariant action we show there is a
non-trivial,  conserved  current $J^\mu$.  We then detail this
  for Weyl and Palatini quadratic gravity. This information was used in
 the text, Section~\ref{MG}, eq.(\ref{current}) and also in the Palatini case, see text after eq.(\ref{opop}).

 Consider  a Weyl gauge transformation:
%\medskip
\bea
\hat g_{\mu\nu}=\Sigma\,g_{\mu\nu},\qquad
\hat\phi=\Sigma^{-1/2} \phi,\qquad
\hat\omega_\mu=\omega_\mu-\frac{1}{\alpha} \partial_\mu \ln\Sigma,
\eea
%\medskip\noindent
where $\phi$ is some scalar field.
For an infinitesimal transformation $\delta \Sigma$ 
\medskip
\bea\label{var}
\delta \hat g_{\mu\nu}=\delta(\ln\Sigma)\,\hat g_{\mu\nu},\quad
\delta\hat\phi=-\frac12\,\delta(\ln\Sigma)\,\hat \phi,\quad
\delta\hat\omega_\mu=-\frac{1}{\alpha}\,\delta\, \partial_\mu\, \ln\Sigma
% -\frac{1}{\alpha}\,\partial_\mu\,\delta\ln\Sigma.
=-\frac{1}{\alpha}\partial_\mu\, \delta\ln\Sigma.
\eea

\medskip
Consider a Weyl gauge invariant {\it total} action which we write  as a  sum $S_g+S$, where
$S_g$ is the Weyl gauge field kinetic term while $S$
 is the {\it remaining}  part of the  action that can depend
on $\omega_\mu$ but not on $F_{\mu\nu}$, hence:
\be
S_g=-\frac14\int \sqrt{g}\, F_{\mu\nu}^2,
\qquad
S=\int \sqrt{g} \,L
\ee
$S_g$ and $S$ are each Weyl gauge invariant.
Under (\ref{var}) 
\medskip
\bea\label{J3}
\delta S=\int \sqrt{\hat g}\,
\Big[
-\frac{1}{2}\,\theta^{\mu\nu}\,\delta \hat g_{\mu\nu} + J^\mu\,\delta\hat \omega_\mu +
\frac{\delta L}{\delta \hat\phi} \,\delta\hat \phi\Big],
\eea

\medskip\noindent
where $\theta^{\mu\nu}$ and $J^\mu$ denote the ``stress-energy''  tensor associated to $S$ and 
the Weyl gauge symmetry  current, respectively.
The last term in $\delta S$ vanishes by the equation of motion for $\phi$. 
Since $S$ is Weyl gauge invariant ($\delta S=0$) and using (\ref{var}),
then  (after removing the ``hat'' notation):
%\smallskip
\bea
0=\delta S\!\!\!&\!\!\equiv\!\! &\!\!\!
\int\!\! \sqrt{g}\Big[
-\frac{1}{2} \theta^{\mu\nu} g_{\mu\nu}
\,\delta \ln\Sigma -\frac{1}{\alpha}J^\mu\,\partial_\mu \delta\ln\Sigma\Big]
\!=\!\int\!\!\sqrt{g} \Big[-\frac12\,\theta^\mu_\mu
+\frac{1}{\alpha} \nabla_\mu J^\mu\Big]\delta\ln\Sigma.\quad\,\,\,
\eea

\medskip\noindent
where $\sqrt{g}\nabla_\mu J^\mu=\partial_\mu (J^\mu \sqrt{g})$ was used.
Therefore, for a Weyl gauge invariant action:
\bea\label{T}
\theta_\mu^\mu=\frac{2}{\alpha}\nabla_\mu J^\mu.
\eea
Finally, from the total action $S+S_g$ one can easily write the equation of motion for $\omega_\mu$:
\bea
J^\mu + \nabla_\sigma F^{\sigma\mu}=0
\eea
Next,  multiply  this equation by $\sqrt{g}$ and  apply $\partial_\mu$ on it,
then use the antisymmetry of $F^{\sigma\mu}$ to find that $\nabla_\mu J^\mu=0$.
Thus, there is a conserved
current, as discussed in the text. Finally, using (\ref{T}), then
onshell
\bea
\theta_\mu^\mu=0,
\eea as expected for a theory with this classical symmetry.

\bigskip\bigskip\noindent
$\bullet$
Let us now detail the above results for our cases of Weyl and Palatini quadratic gravities
that are Weyl gauge invariant, discussed in Section~\ref{MG}.
Consider our initial action in Weyl non-metric case eq.(\ref{alt2})
which is similar for Palatini case, eq.(\ref{opop}). Therefore our $S$ becomes
%\smallskip
\bea\label{SSS}
S\equiv \int \sqrt{g}\,L=
\int \sqrt{g}\, \Big\{
\frac{-1}{12\xi^2}
\,\phi^2\,\Big[ R-3\alpha \nabla_\mu\omega^\mu
-\frac32\,\alpha^2\,\omega_\mu\omega^\mu\Big]
-\frac{\phi^4}{4!\,\xi^2}
%-\frac{1}{4}\,F_{\mu\nu}^2
\Big\}
\eea

\medskip\noindent
Using  (\ref{J3}), we find a current 
\bea\label{jjj}
J_\mu=\frac{1}{\sqrt{g}} \frac{\delta L}{\delta \omega^{\mu}}=
\frac{1}{\sqrt{g}} \frac{\partial L}{\partial \omega^{\mu}}=
-\frac{\alpha}{4\xi^2}\,(\partial_\mu-\alpha\,\omega_\mu)\phi^2.
\eea

\medskip\noindent
The {\it total} action  $S+S_g$ is identical to  action  (\ref{alt2}) in the text and gives 
the following equation of motion for $\omega_\mu$ 
\bea\label{eom}
\sqrt{g}\,\Big\{\,
\frac{\alpha^2}{4\,\xi^2}\phi^2 \,\omega^\rho -\frac{\alpha}{4\xi^2}\,\nabla^\rho \phi^2
+\nabla_\sigma F^{\sigma\rho}\Big\}=0,
\eea

\medskip\noindent
This equation is Weyl gauge invariant, as expected (since  the action  is invariant).
Applying $\partial_\rho$ on the last equation, using that
$\sqrt{g}\,\nabla_\sigma F^{\sigma\mu}= \partial_\sigma\, (\sqrt{g}\,F^{\sigma\mu})$
and the antisymmetry of $F^{\sigma\rho}$, we then find
\bea\label{div}
\nabla_\mu J^\mu=0.
\eea

\medskip\noindent
Therefore, there exists a non-trivial, conserved current. This result  was used in  Weyl case,
eq.(\ref{current}), and in the Palatini case in the text after eq.(\ref{opop}).

Let us now check explicitly eq.(\ref{T}) for our case. 
From action (\ref{SSS}) one has the trace:
\medskip
\be\label{tr}
g^{\mu\nu}\frac{\delta L}{\delta g^{\mu\nu}}=
\frac{\sqrt{g}}{12\,\xi^2}
\,\Big[ \phi^2\Big( R - \frac32 \alpha^2\,\omega_\mu\omega^\mu
- 3\alpha \,\nabla_\rho\omega^\rho\Big)
+\phi^4
- 3 \Box\phi^2 + 3 \alpha\nabla^\rho (\omega_\rho\phi^2)
\Big]=0.
\ee

\medskip\noindent
This vanishes by the equation of motion for $g^{\mu\nu}$.
This result is actually valid  for the total action $S+S_g=\int \sqrt{g}\,
\big[L-(1/4) F_{\mu\nu}^2\big]$  since
the contribution  to the trace by the (conformal) gauge kinetic term $F_{\mu\nu}^2 \sqrt{g}$
is vanishing. Finally, from the equation of motion of $\phi$
(which is also Weyl gauge invariant) one finds, after multiplying it by $\phi$
\medskip
\bea\label{S}
\frac{\sqrt{g}}{12\,\xi^2} \,\Big[
 \phi^2\,(R- 3 \alpha \nabla_\rho\omega^\rho
-\frac32\,\alpha^2\omega_\rho\omega^\rho)+\phi^4\Big]=0,
\eea

\medskip\noindent
Eq.(\ref{S}) is just another form of equation  $\phi^2=-\tilde R$ which we already know
from the ``linearisation'' of the $\tilde R^2$ term, described in the text.
Using eq.(\ref{S}) in eq.(\ref{tr})  we find
\bea\label{opi}
\frac{2}{\sqrt{g}} g^{\mu\nu} \frac{\delta L}{\delta g^{\mu\nu}}=
\frac{-1}{2\,\xi^2} \,\nabla^\rho \Big[( \nabla_\rho - \alpha \omega_\rho)\phi^2\Big] =0.
\eea
The last equation also shows that $\phi$ which was ``extracted'' from the
$\tilde R^2$ term, is indeed a dynamical field, as discussed in the text.
This equation gives again  the current conservation (also found earlier directly  from (\ref{eom})).
From (\ref{J3}),  (\ref{opi}) we have
\medskip
\bea
\theta_\mu^\mu=\Big(\frac{2}{\alpha}\Big)\,\Big(\frac{-\alpha}{4 \xi^2}\Big)
\,\nabla^\rho \,\Big[(\nabla_\rho-\alpha\,\omega_\rho)\phi^2\Big]
=0
\eea

\medskip\noindent
in agreement with general result (\ref{T}) and also   with (\ref{jjj}), (\ref{eom}).
The analysis for  the Palatini case eq.(\ref{opop}) is very similar (with $\omega_\mu\ra v_\mu$).
For more  details, see also \cite{Ghilen1}, \cite{SMW} (Appendix B) and \cite{Palatini1}.

\bigskip
\subsection*{B. Parallel transport in Weyl and Palatini cases}\label{apb}
\def\theequation{B-\arabic{equation}}
\def\thesubsection{B}
\setcounter{equation}{0}
\def\thefigure{B-\arabic{figure}}
\def\thelabel{B}

We present here a brief review of the parallel transport of a vector in
 Weyl and Palatini geometries, discussed in Section~\ref{38}:

\bigskip\noindent
{\bf $\bullet$ Weyl case:\,\,\,}
Weyl geometry\footnote{
  For a brief introduction to Weyl conformal geometry  see Appendix A in \cite{SMW}.}
is represented by  classes of equivalence of $(g_{\mu\nu}, \w_\mu)$ related by (\ref{a3}).
Scalars $\phi$ and fermions $\psi$ transform under (\ref{a3})
as shown in (\ref{a4}) below:
\bea\label{a3}
&&\qquad \hat g_{\mu\nu}=\Sigma^q \,g_{\mu\nu}, \quad
\sqrt{\hat g}=\Sigma^{2\,q}\sqrt{g},\quad
\hat\w_\mu=\w_\mu-\frac{1}{\alpha}\,\partial_\mu\ln\Sigma
\\[5pt]
&&\qquad\hat\phi=\Sigma^{-q/2}\,\phi,\quad
\hat\psi=\Sigma^{-3 q/4} \psi\quad
\label{a4}
\eea

\medskip\noindent
where, to be more general, we now allow an arbitrary Weyl charge $q$ for the metric
(one usually sets $q=1$, as done so far in this work).
Since the metric has  charge $q$ the tetrad $e_\mu^a$ then has charge $q/2$ which is used below.
The gauge covariant derivative of the scalar $\phi$
transforms  just like the scalar itself and equals
$D_\mu \phi=\big[\partial_\mu -(q/2) \,\alpha \,\w_\mu\big]\phi$.

Weyl geometry has vectorial non-metricity
\bea\label{wg}
\tilde\nabla_\mu g_{\alpha\beta}=-\alpha\,q\, \w_\mu \,g_{\alpha\beta},
\eea
where $\tilde\nabla$ is defined by the Weyl connection $\tilde\Gamma$
\bea\label{sdsd}
\tilde\nabla_\mu g_{\alpha\beta}=\partial_\mu g_{\alpha\beta}
-\tilde\Gamma_{\alpha\mu}^\rho g_{\rho\beta}
-\tilde\Gamma_{\beta\mu}^\rho g_{\rho\alpha}.
\eea

\medskip\noindent
Eq.(\ref{wg})  may be written in a ``metric'' format
\bea\label{wq1}
\tilde\nabla^\prime _\mu g_{\alpha\beta}=0,
\qquad \tilde\nabla^\prime\equiv \tilde\nabla\Big\vert_{\partial_\mu\ra\partial_\mu +
  \alpha\, q\, \w_\mu}.
\eea
Therefore the Weyl connection $\tilde\Gamma$ is found from the
Levi-Civita connection ($\Gamma$) in which
one makes the same substitution: $\tilde\Gamma=\Gamma\vert_{\partial_\lambda\ra\partial_\lambda +
  \alpha\, q\, \w_\lambda}$, or by ``standard''
calculation as for the Levi-Civita connection in Riemannian case.
Either way, one finds for a symmetric connection
\bea\label{tGammap}
\tilde \Gamma_{\mu\nu}^\lambda=
\Gamma_{\mu\nu}^\lambda+(q/2)\,\q \,\Big[\delta_\mu^\lambda\,\, \w_\nu +\delta_\nu^\lambda\,\, \w_\mu
- g_{\mu\nu} \,\w^\lambda\Big].
\eea

\medskip\noindent
Consider now  a vector $u^\mu$ of some Weyl charge ($z_u/2$):
\bea
 \hat u^\mu =\Sigma^{z_u/2} u^\mu
\eea
The parallel transport of a constant vector  (in a Weyl-covariant sense) is defined by
\bea\label{pt}
\frac{D\, u^\mu}{d\tau}=0, \quad \text{where}\quad
D\equiv dx^\lambda\,D_\lambda,\quad
D_\lambda\, u^\mu= \tilde\nabla_\lambda u^\mu \Big\vert_{
 \partial_\lambda\ra \partial_\lambda+(z_u/2)\,\alpha\,\w_\lambda},
\eea
with 
\bea
\tilde\nabla_\lambda u^\mu=
\partial_\lambda u^\mu
% + (z/2)\,\alpha\,\w_\beta \,u^\mu
+\tilde\Gamma_{\lambda\rho}^\mu\,u^\rho,
\eea
and $x=x(\tau)$. Then from (\ref{pt}) the ``standard'' differential variation of the vector is
\bea\label{A10}
d\, u^\mu=
-d x^\lambda \,\Big[ (z_u/2)\, \alpha\,\w_\lambda \,u^\mu +\tilde\Gamma_{\lambda\rho}^\mu\,u^\rho\Big],
\quad\text{where}\quad
d\, u^\mu\equiv dx^\lambda \,\partial_\lambda u^\mu.
\eea
Then under the  parallel transport,
the product $\langle u,t\rangle=u^\mu\,t^\nu \, g_{\mu\nu}$ of vectors $u, t$ changes as
\bea\label{prod}
d\langle u,t\rangle
=dx^\lambda \,\Big[\tilde\nabla_\lambda g_{\mu\nu}- \alpha\,\w_\lambda\, g_{\mu\nu} (z_u+z_t)/2\Big]
u^\mu t^\nu.
\eea
This can immediately be integrated along a given path $\gamma(\tau)$.

Using non-metricity (\ref{wg}), then  the norm $\vert u\vert$ of the vector $u^\mu$ varies according to
\bea\label{sss}
d \vert u\vert^2= dx^\lambda \vert u\vert^2\,\w_\lambda\,(-\alpha)\,(q+z_u),
\eea
or, integrating this along a path $\gamma(\tau)$:
\bea\label{norm}
\vert u\vert^2=\vert u_0\vert^2\,e^{-\alpha\, (q+z_u)\,\int_\gamma \w_\lambda dx^\lambda}.
\eea

The integral and thus  the norm of a vector are in general path-dependent.

If  $\w_\mu$ is an exact one-form and the path is closed the integral
vanishes and the norm is invariant; the norm is then independent of path
between any  two points. This is the case of Weyl integrable  geometry.

Note that if a tangent space vector $u^a=e^a_\mu u^\mu$ has a
vanishing charge, i.e. is invariant, then $q/2+z_u/2=0$, since $e^a_\mu$ has
charge $q/2$ while $u^\mu$ has charge $z_u/2$. Then for this case  also the  norm
  itself  is invariant $\vert u\vert =\vert u_0\vert$,
since  the exponent in (\ref{norm}) vanishes ($q+z_u=0$),
under parallel transport for any  $\gamma$,
even though the theory has non-metricity  $\tilde\nabla_\mu g_{\alpha\beta}\not=0$!

Finally, the ratio of the norms of two vectors of same but otherwise {\it arbitrary} Weyl weight
is also invariant under parallel transport on the same $\gamma$, as seen by using (\ref{sss})
\bea\label{b14}
d\,\ln \frac{\vert u\vert^2}{\vert t\vert^2}=(-\alpha)\,
(z_u -z_t)\,\w_\lambda \,dx^\lambda
\eea
which  vanishes if  $z_u=z_t$, result used in Section~\ref{38}.
This is of interest since for all physical purposes,
the relative length  should be invariant under parallel transport
(physics being independent of the units of length).

\vspace{0.4cm}
\bigskip\noindent
$\bullet$ {\bf Palatini case:} In the Palatini case the non-metricity is found  
from action (\ref{L2p}) or equivalent (\ref{opop})
by solving the equations of motion of the connection $\tilde\Gamma$.
This is a rather technical exercise detailed for this action
in ref.\cite{Palatini1} (section 3.1). One finds
\medskip
\bea\label{A15}
\tilde\nabla_\lambda (\phi^2 g_{\mu\nu})=(-2) (g_{\mu\nu}\,V_\lambda -g_{\mu\lambda}\,V_\nu -g_{\nu\lambda} \,V_\mu)\phi^2
\eea
 where
\bea
V_\lambda= v_\lambda -\partial_\lambda\ln\phi^2
\eea

\medskip\noindent
and where $v_\lambda$ is the Weyl gauge boson in the Palatini case.
From this result, one finds  the connection  \cite{Palatini1}
\medskip
\bea
\tilde \Gamma_{\mu\nu}^\alpha&=&\Gamma_{\mu\nu}^\alpha(g)+(1/2) \,(\delta_\nu^\alpha\,\partial_\mu
+\delta_\mu^\alpha\,\partial_\nu-g^{\alpha\lambda}\,g_{\mu\nu} \,\partial_\lambda)\ln\phi^2
\nonumber\\[5pt]
&-& (3 g_{\mu\nu} V_\lambda -g_{\nu\lambda}\,V_\mu -g_{\lambda_\mu}\,V_\nu)\,g^{\lambda\alpha}
\eea

\medskip\noindent
After $\phi$ acquires a vev $\langle\phi\rangle$, (\ref{A15}) becomes
\bea\label{pg}
\tilde\nabla_\lambda g_{\mu\nu}=
(-2) \alpha \,q\,(g_{\mu\nu}\,v_\lambda-g_{\mu\lambda} v_\nu - g_{\nu\lambda}\,v_\mu).
\eea
to be compared to the Weyl case, eq.(\ref{wg}).
With this result,  the change of the norm of a vector 
under the parallel transport can  be computed from (\ref{prod}) in which one is
using  the Palatini non-metricity (with $\omega_\lambda\ra v_\lambda$). One finds
\bea
d\,\vert u\vert^2=\alpha \,dx^\lambda\,\Big[ - v_\lambda 
(2 q+z_u)\,\vert u\vert^2 + 4 \, q\, u_\lambda (u_\beta \, v^\beta)\Big].
\eea
From this one also finds the change of the ratio of the norms of two vectors is non-zero
\medskip
\bea\label{ratioP}
d\,\ln\frac{\vert u\vert^2}{\vert v\vert^2}=
(-\alpha)\Big[
 (z_u-z_t)\,v_\lambda -4 \,q\,v^\beta\ \Big(\frac{u_\lambda u_\beta}{\vert u\vert^2} - 
\frac{t_\lambda t_\beta}{\vert t\vert^2}\Big)\Big]\, dx^\lambda,
\eea

\medskip\noindent
Unlike in  Weyl geometry eq.(\ref{b14}),  the ratio of the norms of $u$ and $t$
changes under parallel transport even if they have the same Weyl charge $z_u=z_t$.
This result was used in Section~\ref{38}.

Finally,  in a most general case, note that in the Palatini quadratic gravity there are
more quadratic operators in curvature that can be present in action (\ref{L2p})
and this makes the analysis much more difficult in such case.
A list of the quadratic operators (in curvature) that can be present
is found in \cite{Borunda}, see also \cite{ST3}.
In such case there are additional states propagated by these (higher derivative)
operators, other than  $\phi$ (propagated by $\tilde R^2$); some of these  may even be
ghost-like; in such case it is unclear if one can still solve algebraically
 the second-order differential  equations of motion for the
 Palatini connection, since these equations acquire new terms with new indices
 structure and new states present.

\bigskip\bigskip
\noindent
{\bf Acknowledgement:}\,

\medskip\noindent
This work was supported by a grant of the Romanian Ministry
of Education and Research, CNCS-UEFISCDI, project number PN-III-P4-ID-PCE-2020-2255 (PNCDI~III).

\end{document}